\documentclass[conference]{IEEEtran}
\IEEEoverridecommandlockouts
\usepackage{amsmath,amssymb,amsfonts}
\usepackage{algpseudocode}
\usepackage{graphicx}
\usepackage{textcomp}
\usepackage{booktabs}
\usepackage{enumitem}
\usepackage{subfigure}
\usepackage[colorlinks,bookmarksopen,bookmarksnumbered,citecolor=red,urlcolor=red]{hyperref}

\def\BibTeX{{\rm B\kern-.05em{\sc i\kern-.025em b}\kern-.08em
		T\kern-.1667em\lower.7ex\hbox{E}\kern-.125emX}}
	
\begin{document}
\newcommand{\tabincell}[2]{\begin{tabular}{@{}#1@{}}#2\end{tabular}}
%

%

%
\ifCLASSINFOpdf
\else
\fi
\hyphenation{op-tical net-works semi-conduc-tor}

\newenvironment{figurehere}
  {\def\@captype{figure}}
  {}
\makeatother

	\title{Joint Communication and Computational Resource Allocation for QoE-driven Point Cloud Video Streaming}
	

\author{
\IEEEauthorblockN{Jie Li$^\dagger$, Cong Zhang$^\dagger$, Zhi Liu$^\ddagger$, Wei Sun$^\mp$, Qiyue Li$^\mp$}
\IEEEauthorblockA{$^\dagger$School of Computer and Information, Hefei University of Technology, China\\
$^\ddagger$Department of Mathematical and Systems Engineering, Shizuoka University, Japan \\
$^\mp$ School of Electrical Engineering and Automation, Hefei University of Technology Hefei, China\\
Email: \{lijie@hfut.edu.cn, zhangcong@mail.hfut.edu.cn, liu@shizuoka.ac.jp  \\wsun@hfut.edu.cn, liqiyue@mail.ustc.edu.cn\}
}
}	
	

\maketitle
	
\begin{abstract}

Point cloud video is the most popular representation of hologram, which is the medium to precedent natural content in VR/AR/MR and is expected to be the next generation video. Point cloud video system provides users immersive viewing experience with six degrees of freedom (6DoF) and has wide applications in many fields such as online education and entertainment. To further enhance these applications, point cloud video streaming is in critical demand. The inherent challenges lie in the large size by the necessity of recording the three-dimensional coordinates besides color information, and the associated high computation complexity of encoding/decoding.
To this end, this paper proposes a communication and computational resource allocation scheme for QoE-driven point cloud video streaming. In particular, with the goal to maximize the defined QoE by selecting
proper quality levels (uncompressed tiles at different quality levels are also considered) for each partitioned point cloud video tile, we formulate this into an optimization problem under the limited communication and computational resources constraints and propose a scheme to solve it. Extensive simulations are conducted and the simulation results show the superior performance of the proposed scheme over the existing schemes.

\end{abstract}
	

%

\section{Introduction} \label{sec_intro}
With the rise of the immersive video, technologies such as Virtual Reality (VR) and Augmented Reality (AR) are increasingly favored by users. Currently 360 degree videos \cite{liu2018jet,guo2018optimal,8957509} are widely used because 360 degree video can directly use the standard video compression technology such as HEVC for encoding. However, 360 degree video can only achieve two degrees of freedom (2DoF), and this limits the immersive viewing experience.
Thus, hologram video, which can support six degrees of freedom (6DoF) and can give users a real immersive experience, becomes a research hotspot and draws more and more attentions \cite{8068913}.

However, higher degree of freedom also means a much larger amount of data to be transmitted with stringent playback deadline, thus the wireless streaming of hologram videos brings more challenges. Nowadays, light field and point cloud are two commonly used technologies to represent the hologram video.
Due to the difficulty of obtaining the full plenoptic function for scene, light field technology is less popular in practice \cite{8022901}, and point cloud is more proper for hologram video applications. Note that the point cloud is composed of quantities of points in three-dimensional space, and each point contains RGB attributes. Due to that the point cloud needs to store three dimensions of information, the amount of data is very large compared to ordinary video \cite{8019426,hu2019orchestrating}. For example, when the number of 3D points is around 2,800,000, the amount of traffic without any compression is approximately 78 Mbits for a single frame, which means that the bandwidth requirement of hologram video streaming is about 2300Mbps when frame rate is 30 frames per second. This number will be even lager when streaming high quality point cloud video. Therefore, larger bandwidth requirement is one challenging research issue for the transmission of hologram video.

Currently, there are limited researches about hologram video streaming, and state-of-the-art point cloud researches mainly focus on the point cloud compression. In particular, Hu, et al. propose a novel point cloud compression method for attributes, based on geometric clustering and normal weighted graph Fourier transform \cite{8462684}.
Mohammad, et al. propose spatially sub-sample point clouds in the 3D space method to reduce data amount and combine DASH protocol to make the point cloud video transmission adaptive \cite{Hosseini:2018:DAP:3210424.3210429}. There are also public point cloud compression tools available, such as Google's Draco \cite{Draco} and point cloud compression (PCC) by MPEG-I \cite{8571288}. However, in our experiment tests, the running time of encoding and decoding is very large, although they support  multi-core, multi-thread operation. For example, the time for a typical computer with i7 processor to decode 30 frames of a popular point cloud video sequence \textit{loot} may be at the minute level.

To reduce the bandwidth consumption, we can use the idea of 360 video processing method \cite{liu2018jet,guo2018optimal,8647729} by partitioning the point cloud video into 3D tiles and only transmitting the tiles within user's FoV. This reduces the number of tiles transmitted and hence saves the precious bandwidth.
In particular, Jounsup, et al propose a volumetric media transmission scheme that cuts the point cloud video into tiles spatially, while different tiles have various quality levels, culling them or reducing their level of quality depending on their relation to the user’s view frustum and distance to the user \cite{8638765}. However, the large encoding and decoding time before playback is not considered in their research. Arunkumar, et al. study the relationship between the decoding time and transmission time in VR video and propose to balance the relationship between the two by partially transferring raw tiles of VR video \cite{Ravichandran:2018:FLL:3241539.3267781}. However, this research does not optimize this process and can not be directly used in point cloud video streaming due to the unique features of point cloud video.

To this end, we study the point cloud video streaming considering both the communication and computational resource. Specifically, the point cloud video is first divided into 3D tiles evenly and each tile is encoded into different quality levels for selection.
Then we optimize the quality level selection (the uncompressed tiles at different quality levels are also considered which do not require any decoding time) to maximize the defined user QoE under the communication and computational resource constraints. This problem is a non-linear integer programming problem and we obtain a sub-optimal solution of this problem.
Extensive simulations are conducted and the simulation results show that the proposed scheme results in higher QoE over baseline schemes.

To the best of our knowledge, this is the first research that jointly considers communication and computational resource allocation for hologram video streaming optimization.
Our contributions are summarized as follows:
\begin{enumerate}
\renewcommand{\labelenumi}{(\theenumi)}
\item We propose a joint communication and computational resources allocation framework for point cloud video streaming, in which computational resources are used to decode the compressed 3D tiles on user's playback device.

\item We propose a QoE evaluation metric for dynamic point cloud video based on user's perspective and point cloud characteristics, including distance between user and the scene and quality level of each tile.
\item We formulate the point cloud video transmission into an optimization problem to maximize the QoE and propose a scheme to solve it.
\item Extensive simulations are conducted and the simulation results show that the proposed scheme results in higher QoE over the baseline schemes.
\end{enumerate}

The remainder of this paper is organized as follows: Section \ref{sec_system} introduces the proposed transmission system, and Section \ref{sec_Problem_formulation} introduces the joint communication and computational resource allocation problem to maximize user's QoE and how we solve this optimization problem. Section \ref{sec_experiment_result} shows the simulation results, and we conclude this paper in Section \ref{sec_conculsion}.

\section{Overview of Point cloud Video Streaming System} \label{sec_system}

In this section, we introduce our point cloud video streaming system and its tile selection module.

\subsection{System Architecture}
\label{sec_system_model}

\begin{figure}
	\centering
	\includegraphics[width=3.5in]{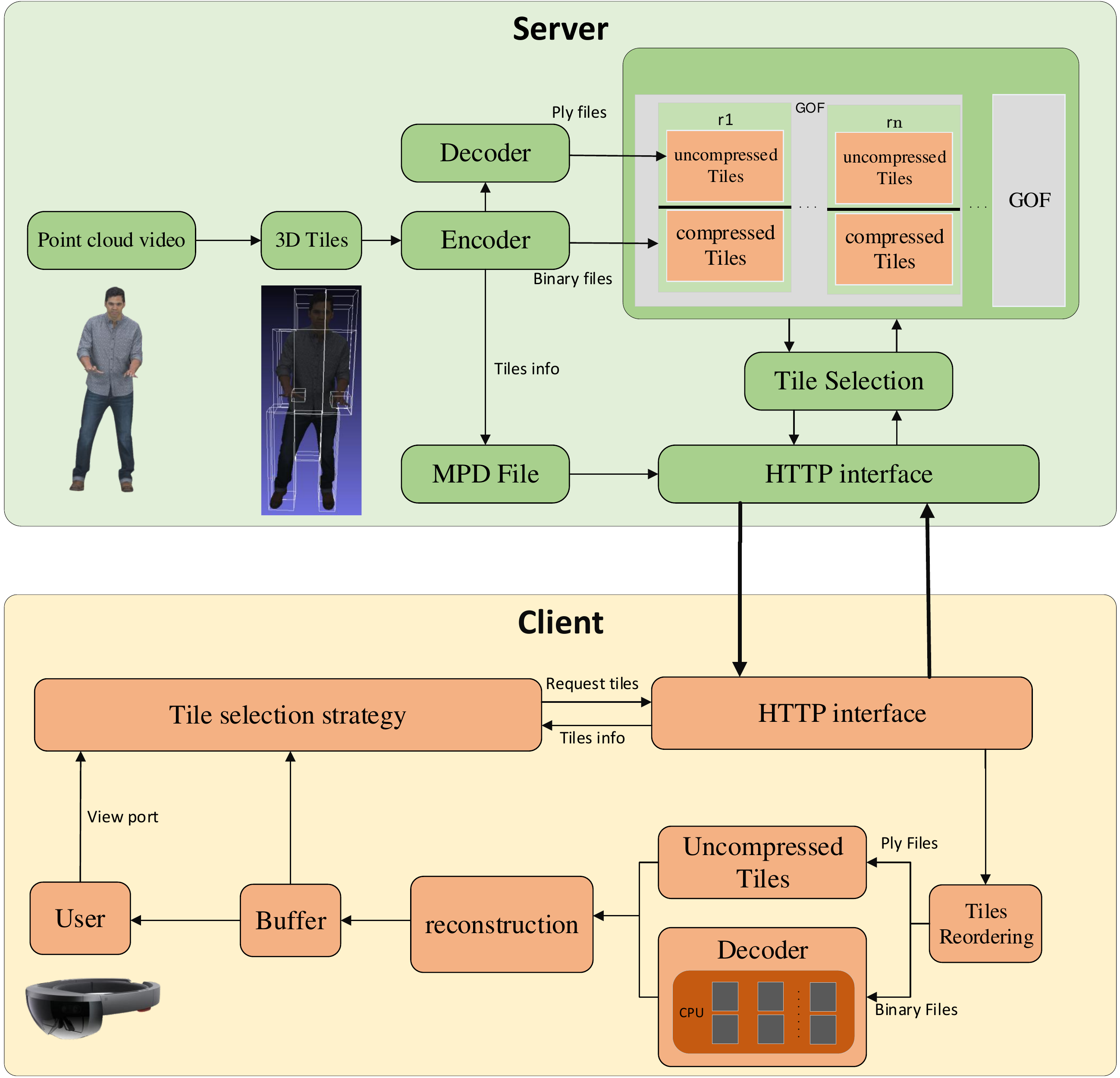}
	\caption{Illustration of the adaptive point cloud video transmission system.}
	\label{fig_system_model}
\end{figure}

As mentioned above, point cloud video cannot be directly transmitted via the wireless network due to its large size. Similar to 360 video streaming, point cloud video is first partitioned into multiple 3D tiles and only the tiles inside user's FoV are transmitted to reduce the bandwidth requirement \cite{8638765}. Besides, these tiles are compressed to further reduce the bandwidth requirement, where decoding requires computational resources of the user's playback device.
With different available computational resources and time-varying network bandwidths, how to optimize the transmission  to achieve a better user's QoE still remains a non-trivial issue. In this paper, we propose a dynamic point cloud adaptive streaming system to select the proper quality level (the raw tiles at different quality levels are also considered which do not require computation power to decode) for each tile inside the FoV and thus conduct the tradeoff between the computational resources and communication resources. 

As shown in Fig.\ref{fig_system_model}, the whole system is composed of two parts: the server side and the client side. The server pre-processes the point cloud video (i.e. partitioning it into 3D tiles) and encodes each tile into different representations at different quality levels with Group of Frame (GOF) as the minimum unit.
The information of all the point cloud video tiles is stored in a Media Presentation Description (MPD) file, similar to MPEG-DASH \cite{MPEG-DASH}, and the server will send corresponding tiles after receiving streaming requests from the client. 
After all the tiles are received and decoded,
the point cloud video will be reconstructed.  Then the reconstructed point cloud video will be sent to the buffer, waiting for playback. To maintain a continuous playback, the buffer cannot be drained.

The core of the client side is the tile selection module. It calculates user's FoV, selects the tiles residing inside the FoV with appropriate quality levels to maximize user's QoE, according to the bandwidth status, buffer status and the available computational resources. Note that the uncompressed tiles at different quality levels are considered in the tile selection, which do not consume computation power for decoding with relatively large source rate. Please also note that considering the decoding complexity is not necessary in traditional video streaming, which also distinguishes the point cloud video streaming from the traditional video streaming.

\subsection{Tile Selection Module} \label{sec_Tile_selection_strategy}

\begin{figure}[htpb]
	\centering
	\includegraphics[width=1in]{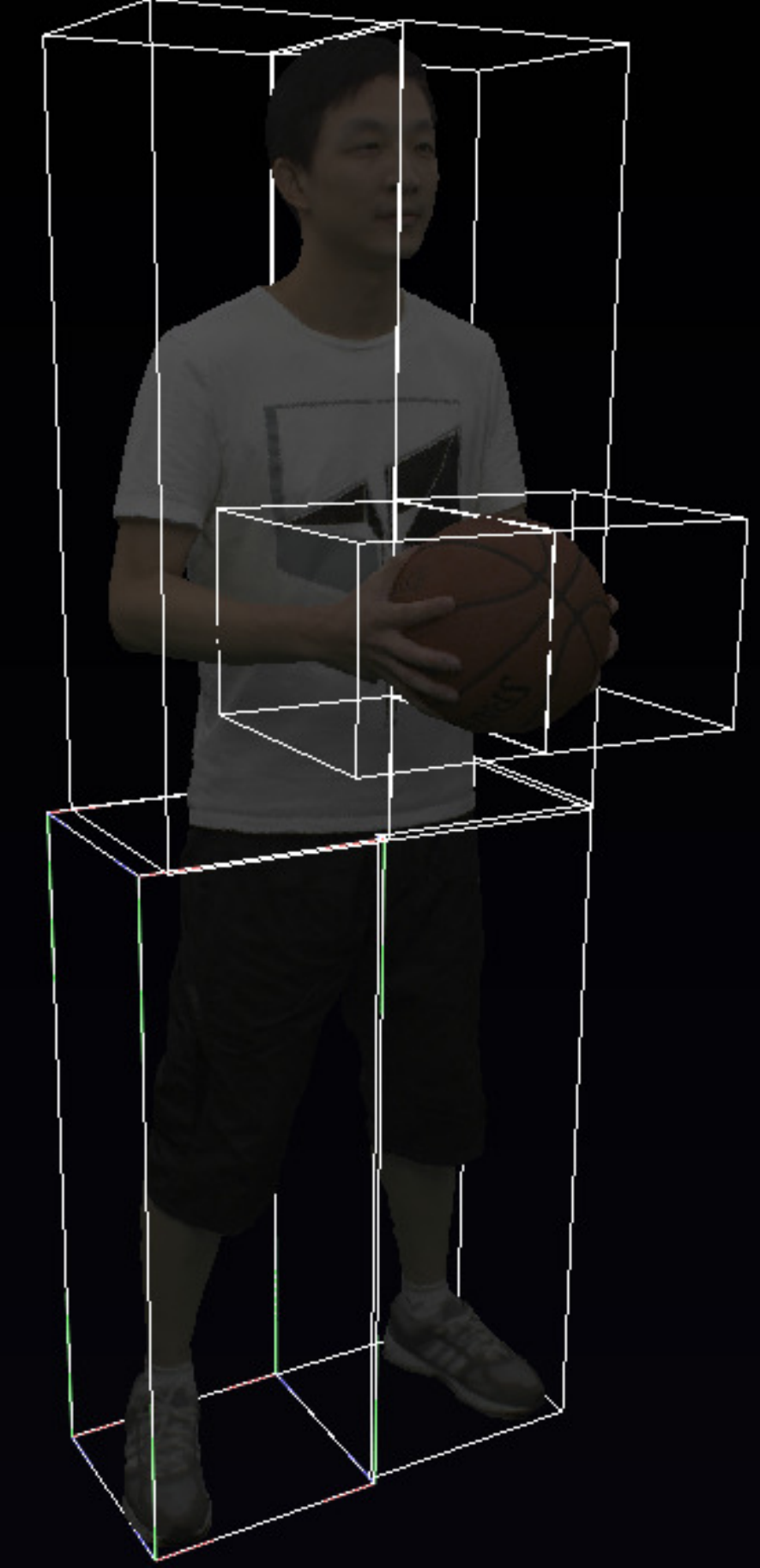}
	\centering
	\includegraphics[width=0.81in]{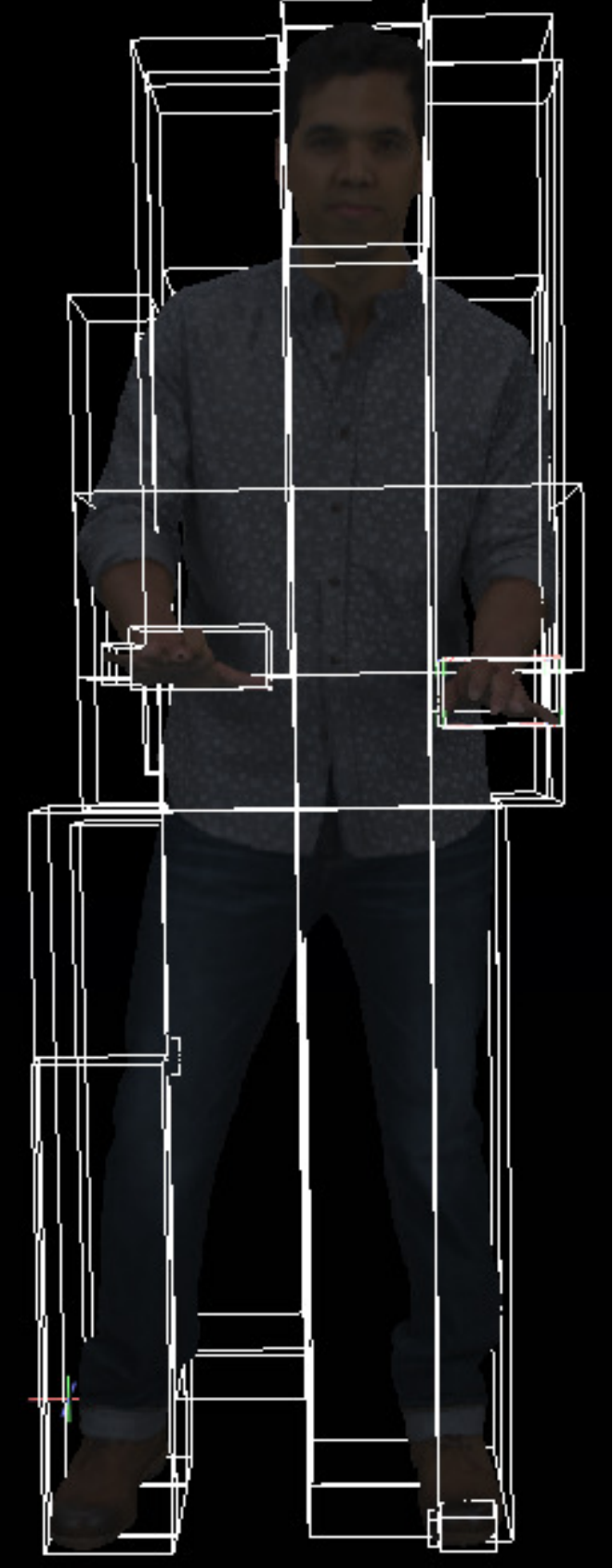}
	\centering
	\includegraphics[width=0.76in]{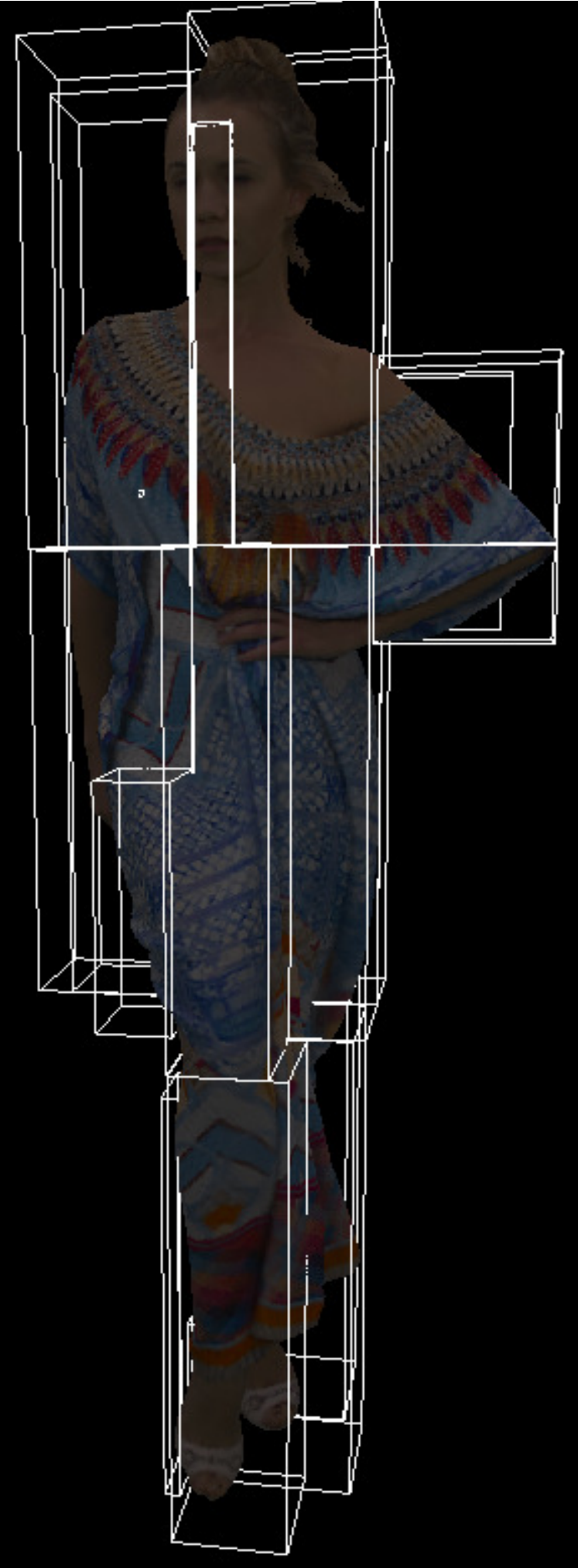}
	\caption{Illustration of the point cloud partitioning. The three sequences are \textit{basketball, loot and longdress}, respectively.}
	\label{fig_tiles_sample}
\end{figure}

To accommodate the transmission of high quality point cloud video, of which the size of compressed video is still large for the existing wireless networks, we borrow the idea of 360 transmission method, i.e. partitioning the point cloud video into 3D tiles evenly and only transmitting the tiles within the user’s FoV. To partition the point cloud video, we first calculate the length, width, and height of the smallest enclosing cuboid that surrounds the point cloud, and determine which side of the cuboid is 'height' according to the situation of the object. For example, when the point cloud is a character, the direction in which the character stands is 'height' of the cuboid. Then we perform $n \times m$ partitioning on the plane perpendicular to the height, and dividing them into $h$ layers in the "height" direction, and finally $n \times m \times h$ tiles are obtained.
Fig.\ref{fig_tiles_sample} shows the point cloud partitioning with three sequences \textit{basketball, loot and longdress}.

With the help of tile partitioning, we can reduce the transmission of non-necessary tiles and further select proper quality level for each tile under the bandwidth and computational resources constraint to optimize the QoE. Note that the decoding from the compressed tiles requires extra computational power (which is much larger than the computational power of decoding a common compressed video) and time, which may limit the usage of the point cloud video compression, while uncompressed tiles do not require the computational resource for decoding with high source rate. How to best utilize the communication and computational resources to provide higher QoE is the focus of this paper.



\section{Problem formulation} \label{sec_Problem_formulation}

\subsection{Computational Resources and Decoding Time} \label{sec_computational_resources}

The decoding time and the required computational resource of point cloud video is related to its total number of points and quality levels \cite{MPEG-ctc}.
Assume the required computational resource to decode one tile with the lowest quality is $C$, we can find a parameter $\partial_{g, k}$, which is related to the number of points and the degree of the compression so that the required computational resource of tile $k$ with quality level $r$ in GOF $g$ can be described as $C_{g,k,r}=\partial_{g,k,r} \times C$, and it can be normalized into CPU frequency to describe the required time to decode the tile. All the consumed computational resource $C_{g,k,r}$ can be regarded as known parameters stored in the MPD file. We assume $CU_{1}$ is the computational resources that one single processor core of user's playback device can provide within one GOF time, and $NC$ is the number of cores, then $CU_{NC}=NC \times \tau_{NC} \times C U_{1}$, where $CU_{NC}$ is the total computational resource available, and $\tau_{NC}$ is the conversion efficiency when the decoding process is running in multi-core and multi-thread mode. Obviously, the consumed computational resource cannot exceed user device's capacity.
Then we can obtain the decoding time $Td_{g}$ for GOF $g$, which can be expressed as follows:
 
\begin{equation}
Td_{g}=\frac{f}{fps} \times \frac{\sum_{k=1}^{M} e_{g,k} \times \sum_{r=1}^{R} C_{g,k,r} \times v_{g,k}\times x_{g,k,r}}{NC \times \tau_{NC} \times CU_{1}}
\end{equation}
 Where $f$ is the number of frames in one GOF and $fps$ is frame rate of the point cloud video.
$e_{g,k}$ is a binary variable, $e_{g,k}=1$ if the compressed version is selected, and $e_{g,k}=0$ means that the uncompressed tile is selected. $M$ is the number of tiles, and $R$ is the number of quality levels available. $v_{g,k}$ is a binary variable calculated through FoV, $v_{g,k}$=1 means the tile is in user's FoV and should be transmitted and $0$ otherwise. $x_{g,k,r}$ is a binary variable, $x_{g,k,r}=1$ means compressed version of tile $k$ with quality level $r$ in GOF $g$ is transmitted, and $0$ otherwise. This equation shows that the decoding time can be determined by the relationship between the computational resources required to decode the point cloud in molecule and the computational resources provided by the device in denominator.

In our system, all the retrieved tiles are stored in a buffer before playback. Let  $Tb_{g}$ denote the current buffer status, which records the frames (measured in time) of point cloud video stored in buffer. To maintain a continuous playback, we have $Tb_{g}>0, \forall g\in [1,G]$, where $G$ presents the total number of GOFs in the video.
Then the dynamics of the playback buffer can be expressed by the following  equations:
\begin{equation}
Tb_{g}=Tb_{g-1}-Tu_{g}+Ti_{g},
\end{equation}
where $Ti_{g}$ is the increased playback time when GOF $g$ enters the buffer, which is a constant value and equals to $\frac{f}{fps}$.
$Tu_{g}$ indicates the time consumed by transmitting all the tiles residing in FOV and decoding the compressed tiles in GOF $g$. Thus $Tu_{g}$ can be expressed as:

\begin{equation}
    Tu_{g}=Ts_{g}+Td_{g}\\
\end{equation}
where
\begin{equation}
\begin{split}
    Ts_{g} = &\frac{\text{$raw\ tiles.size + compressed \ tiles.size$}}{Bw_{g}}\\
   &=\frac{\sum_{k=1}^{M} e_{g, k} \times \sum_{r=1}^{R}binS_{g, k,r}\times v_{g,k} \times x_{g, k, r}}{B w_{g}}\\
    &+\frac{\sum_{k=1}^{M}\left(1-e_{g, k}\right) \times \sum_{r=1}^{R}plyS_{g, k, r} \times v_{g,k} \times x_{g, k, r}}{B w_{g}}
\end{split}
\end{equation}
$binS_{g,k,r}$ indicates the data size of the compressed tile $k$ with quality level $r$ in GOF $g$, $plyS_{g,k,r}$ represents the data size of raw tile $k$ with quality level $r$ ,which is decoded from the compressed file of the corresponding quality level in GOF $g$, and the parameters are all stored in the MPD file. $Bw_g$ is the predicated wireless bandwidth during the time of GOF $g$.

\subsection{The Quality of Experience Model}
\begin{figure}[htpb]
    \centering
    \includegraphics[width=4in]{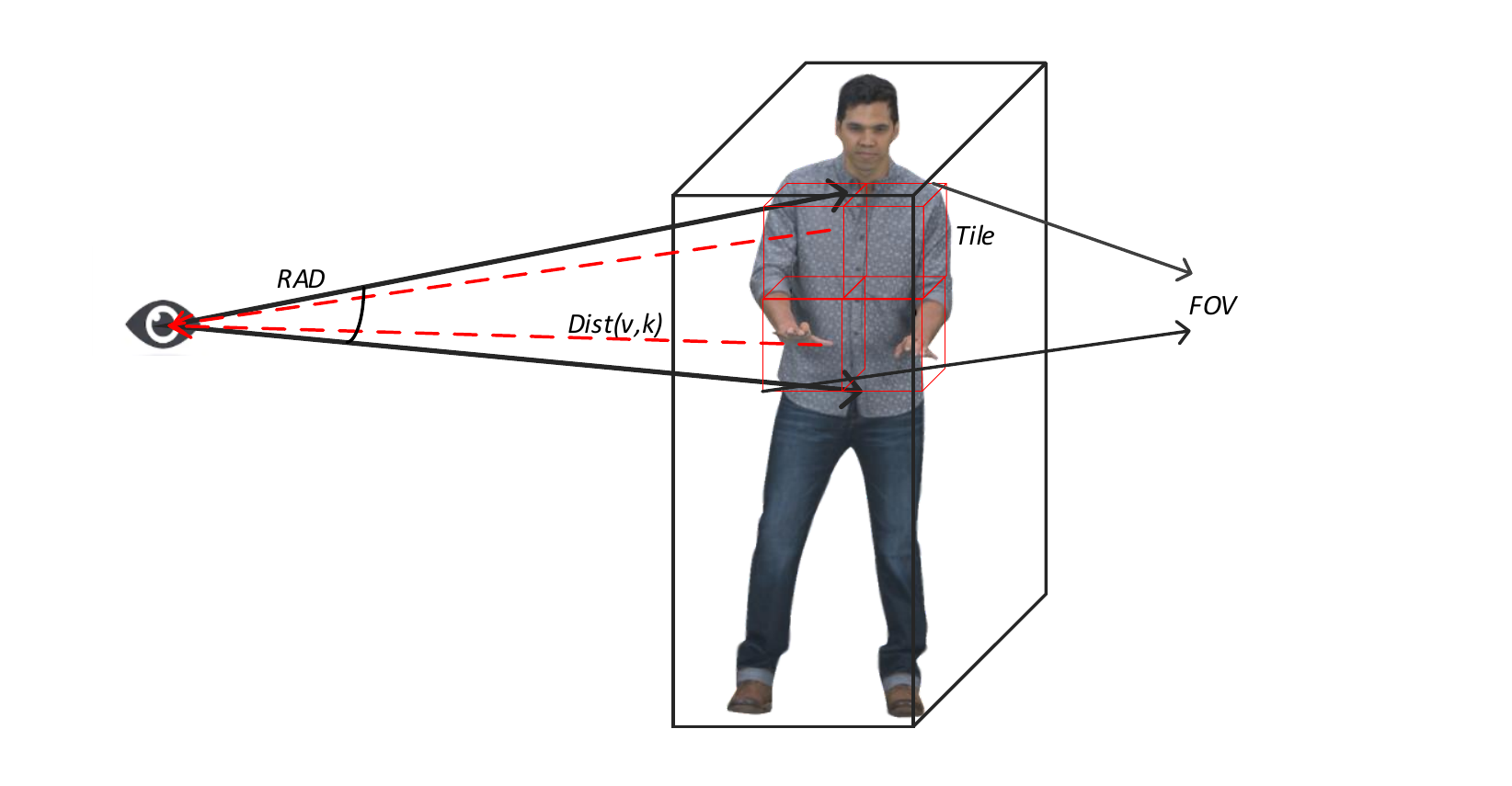}
    \caption{An example of one FoV.}
    \label{fov}
\end{figure}

The point cloud video system can support 6DoF (one FoV is shown in Fig.\ref{fov}). During the playback, different tiles have different distances from the viewer's position, and they have different contributions to the total QoE. We assume the closer the tile is to the user's viewpoint position, the greater its contribution. Besides, for each tile, there are up to $R$ compressed quality levels available, and when different quality level is requested, different QoE will be yielded.

Then for a single tile $k$ in GOF $g$, we can define its QoE contribution as:
\begin{equation}
    Q_{g,k}=\sum_{r=1}^{R} p_{g,k} \times r \times QT_{g,k} \times v_{g,k}\times x_{g, k,r}
\end{equation}
where $p_{g,k}$ and $QT_{g,k}$ are distance weight and quality weight for 3D tile $k$ in GOF $g$, respectively. This QoE model takes the quality of the point cloud video, the proportion of each tile and its degree of influence of the user into consideration.
In this paper, we define $p_{g,k}$ as $
    p_{g, k}=\frac{1}{\operatorname{dist}_{g, k}\left(v_{g}, p o s_{g, k}\right)}$, where $v_{g}$ is the user viewpoint position when watching GOF $g$, and $pos_{g,k}$ is the position of tile $k$ in GOF $g$. $dist_{g,k}$ is a function of the tangential distance from the viewpoint. 
As to the  quality weight $QT_{g,k}$, due to irregularity of the point cloud video, the number of points contained in each tile is different. Thus we can define the quality weight for a 3D tile $k$ as the ratio of number of points in that tile to the total number of points in whole FoV, which can be expressed as $QT_{g,k}=\frac{N_{g,k,R}}{\sum_{k=1}^{M} N_{g,k,R}}$.
The quality of the point cloud frame is generally evaluated by the level of density, which represents the number of points in the unit volume, and the user views each point in the FoV. Then for the whole point cloud video, we can define the QoE as \cite{DBLP:journals/corr/abs-1803-07789}:
\begin{equation}
	\operatorname{QoE}=\log \left( \frac{\sum_{g=1}^{G} \sum_{k=1}^{M} Q_{g,k}}{\sum_{g=1}^{G} \sum_{k=1}^{M} p_{g,k} \times R \times QT_{g,k} \times v_{g,k}}\right)
\end{equation}







\subsection{Objective Function and Solution}
Thus, we formulate the QoE driven communication and computational resource allocation for point cloud video streaming as follows:
\begin{equation}
\begin{split}
   &\max\limits_{x_{g,k,r},e_{g,k}} QoE=\log \left(\frac{\sum_{g=1}^{G} \sum_{k=1}^{M} Q_{g,k}}{\sum_{g=1}^{G} \sum_{k=1}^{M} p_{g,k} \times R \times QT_{g,k} \times v_{g,k}}\right)\\
    &= \log \left(\frac{\sum_{g=1}^{G} \sum_{k=1}^{M} \sum_{r=1}^{R}\left(p_{g,k} \times r \times QT_{g, k}\times v_{g,k} \times x_{g, k, r}\right)}{\sum_{g=1}^{G} \sum_{k=1}^{M} \left( p_{g,k} \times R \times QT_{g,k} \times v_{g,k}\right)}\right)\\
\end{split}
\end{equation}\\
$s.t.$
\begin{equation}\label{equ_x}
\sum_{r=1}^{R} x_{g, k, r}=1, \forall g \in[1, G], k \in\left[1, M\right]
\end{equation}

\begin{equation} \label{equ_constarint_buffer}
Tb_{g}>0, \forall g\in [1, G]
\end{equation}

This is a constrained nonlinear 0-1 programming problem.
Among them, Eq. (\ref{equ_constarint_buffer}) can be calculated as follows:

\begin{equation}
\begin{split}
Tb_{g}&=Tb_{g-1}-Tu_{g}+Ti_{g}\\
&=T b_{g-1}-\left(\frac{\sum_{k=1}^{M}\left[e_{g, k} \times \sum_{r=1}^{R}\left(b i n S_{g, k, r} \times v_{g,k}\times x_{g, k, r}\right)\right]}{Bw_{g}}\right.\\
 &+\frac{\sum_{k=1}^{M}\left[\left(1-e_{g, k}\right) \times \sum_{r=1}^{R}\left(plyS_{g, k, r} \times v_{g,k}\times x_{g,k,r}\right)\right]}{B w_{g}}\\
&\left.+\frac{f}{fps} \times \frac{\sum_{k=1}^{M}\left[e_{g,k} \times \sum_{r=1}^{R}\left(C_{g,k,r} \times v_{g,k}\times x_{g, k,r}\right)\right]}{CU_{NC}}\right)+\frac{f}{fps}
\end{split}
\end{equation}
It can be simplified to   

\begin{equation}
    Tb\left( g\right)-Tb\left( g-1\right)=F\left(x_{g, k, r}\right)
\end{equation}

\begin{equation}
    T b(0)=b>0
\end{equation}

This constraint is a first-order non-homogeneous linear difference equation, $x$, $e$ is a binary vector, and $b$ is the initial buffer size. Then we have
\begin{equation}
    T b(g)=\sum_{i=1}^{g}\left[F\left(x_{i,k,r}\right)\right]+b>0
\end{equation}
The goal of the transmission optimization model we established is to maximize the QoE value. The variables are the quality level $x$ limited by Eq. (\ref{equ_x}) and the transmission form $e$ of each tiles, which presents the compressed or uncompressed tile is transmitted and limited by Eq.(\ref{equ_constarint_buffer}). The problem is a nonlinear integer programming problem. If the variables of the problem is relaxed to a continuous variable, its objective function and constraint are convex functions and convex sets that are easy to solve. Therefore, the paper firstly relaxes the problem, converts the nonlinear integer programming problem problem into nonlinear programming problem, and uses the KKT-condition to find the solution of the nonlinear programming problem. Then the branch-and-bound method is used to solve the integer variable solution of the original problem.

\section{Performance Evaluation} 
\label{sec_experiment_result}

We build a simulation platform to verify the feasibility of our proposed transmission scheme, calculate the different results under different computational resources $CU_{NC}^{i}$ and different communication resources $Bw^{i}$, and then compare with the traditional transmission scheme, that is, the scheme of only transmitting the compressed tiles, in terms of system resource utilization and QoE values. 

Limited to the effectiveness of bandwidth prediction, we set the length of the dynamic point cloud to 2s, with each GOF lasts for 1/3 second. We set $M=2\times 2\times 6$, total quality levels $r=5$ and initial buffer length $b=2s$. The required computational resource with the smallest size among all the tiles is set to $C=1$, and the computational resources required for each tile are obtained according to their densities.
Other simulation parameters are listed in Table \ref{table_group}.
	\begin{table}[h]
	\centering
	\caption{Simulation Parameters Setting}\label{table_group}
	\centerline{}
	\begin{tabular}{c|c|c}
		\hline
		 Group ID	& $NC(Number of Cores)$ &$Bw(Mbps)$	\\\hline
		1	& 2 &54\\
		2	& 2 &72.2\\
	    3	& 2 &104\\
		4   & 4 &54\\
		5   & 4 &72.2\\
		6  &4 &104\\
		7  &6 &54\\
		8  &6 &72.2\\
		9 & 6 &104\\
		\hline
	\end{tabular}
\end{table}


\begin{figure}[htb!]
  \centering
  \subfigure[]{
    \label{fig:subfig:tileform}  
    \includegraphics[scale=0.32]{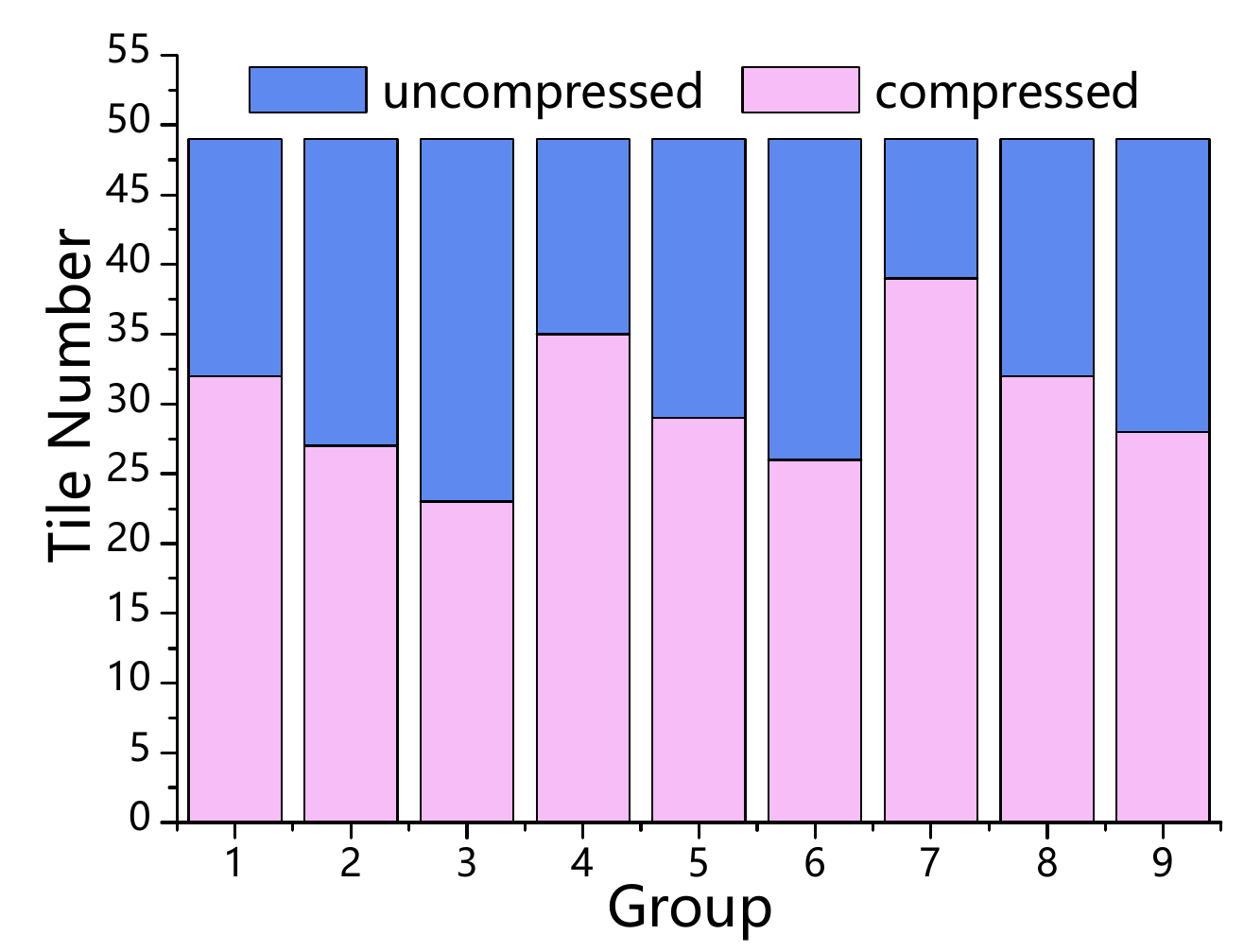}}
    \hspace{-0.12in}
  \subfigure[]{
    \label{fig:subfig:tiler} 
    \includegraphics[scale=0.32]{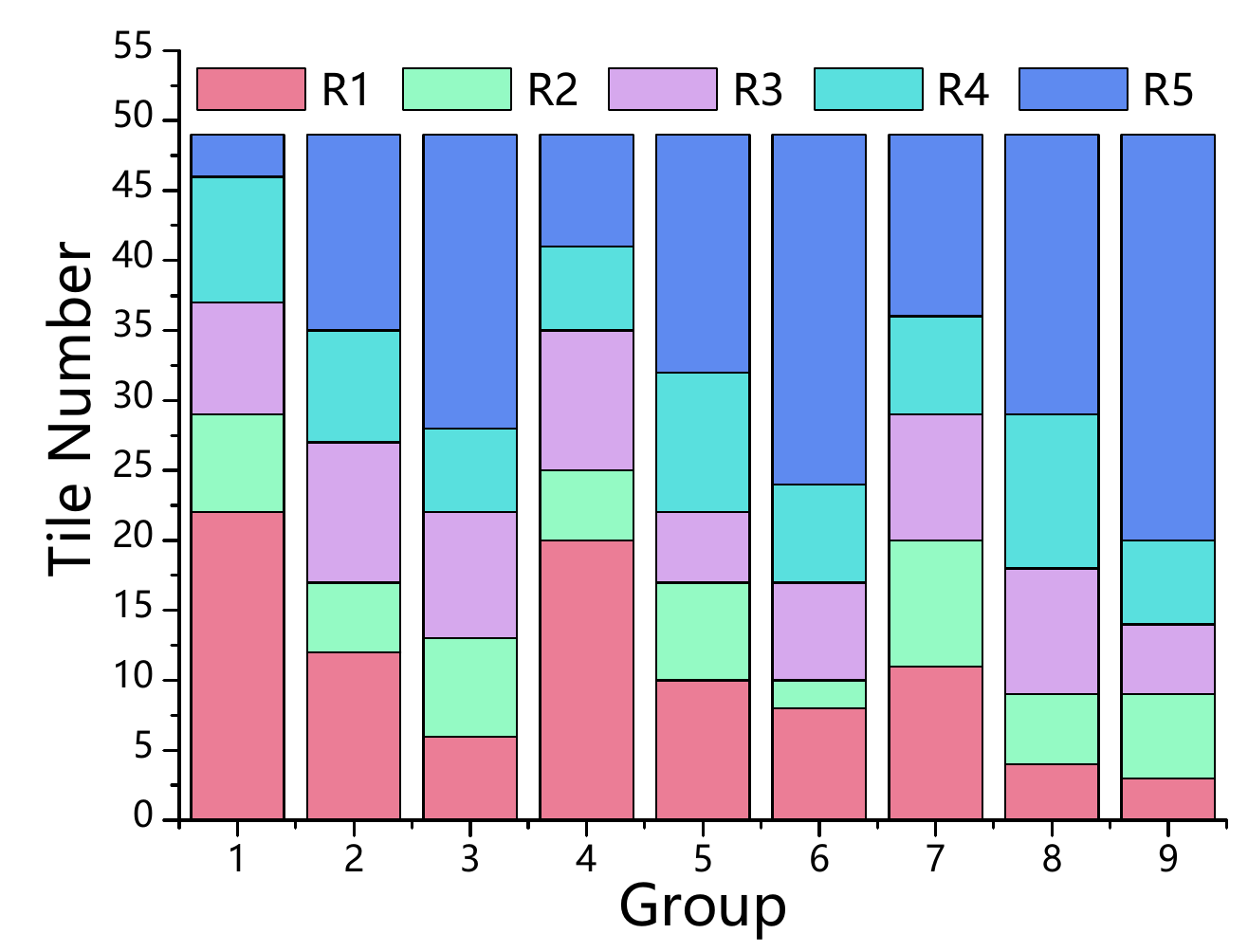}}
  \caption{ (a) Number of tile transmission forms under different schemes    (b) Comparison of the number of tile in different quality under different condition}
  \label{fig_compare}
\end{figure}






Firstly, we study the performance of our proposed system. Fig.\ref{fig:subfig:tileform} shows the number of transmitted uncompressed tiles and compressed tiles under various conditions. We can see that when the computational resources are relatively large (with larger $NC$), the scheme preferentially transmits the compressed tile to improve the resource utilization. When the bandwidth is relatively large, the scheme preferentially transmits the uncompressed tile to make full use of the communication resources, in order to achieve the best system resource utilization.

Fig.\ref{fig:subfig:tiler} illustrates the difference in terms of tile numbers at different quality levels (where R1 is the lowest quality and R5 is the highest quality level) transmitted under different conditions. When the computational and communication resources are low, the system will choose to transmit tiles at a lower bit rate and quality level to reduce resource requirements, and when the system resources are sufficient, high different quality tile will be transmitted to achieve the best quality experience. For example, when comparing group 1, 2, 3, we can find that the number of high quality level tiles becomes larger.

\begin{figure}[htpb]
    \centering
    \includegraphics[width=2.9in]{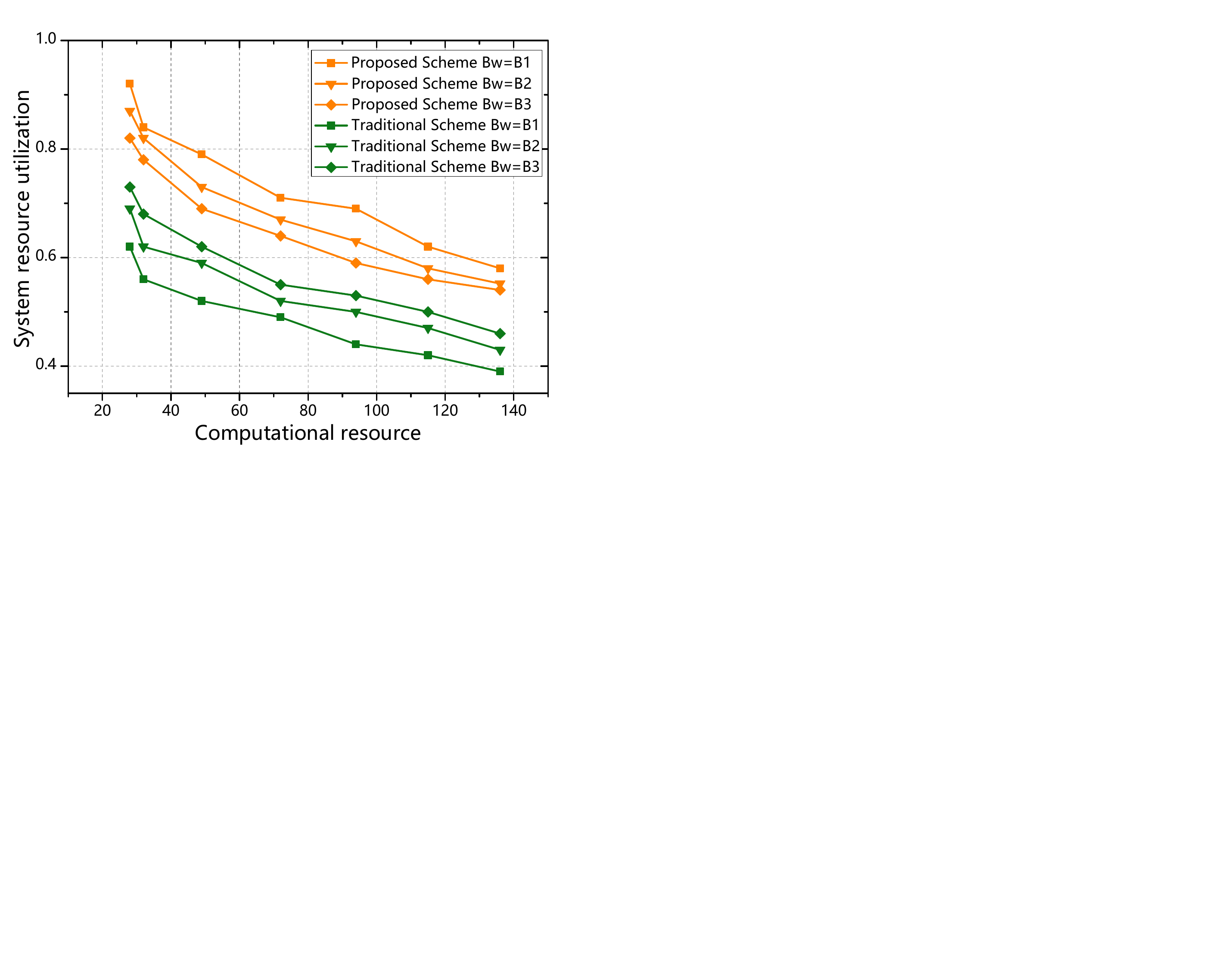}
    \caption{System utilization comparison}
    \label{u_compre}
\end{figure}

Then we compare with traditional hologram transmission system, which delivers only compressed tiles. Three bandwidths are considered, i.e. $B1=54Mbps, B2= 72.2Mbps, B3=104Mbps$ as shown in TABLE \ref{table_group}.
Fig.\ref{u_compre} illustrates system resource utilization expressed by $p=\frac{1}{2}\times(\frac{tiles.C}{ system.C}+\frac{tiles.B}{system.B})$, where $tiles.c$ represents the total computational resources required to decode the tiles. $system.c$ represents the available computational resources, $tiles.B$ represents the total amount of data required for transmission, and $system.B$ represents the communication resources that system can be provided. The x-axis represents the computational resources provided by the device, i.e. the CU mentioned above.
From the simulation results, we can find that our system has a better system resource utilization in comparison to the traditional hologram transmission scheme.

\begin{figure}
    \centering
    \includegraphics[width=2.9in]{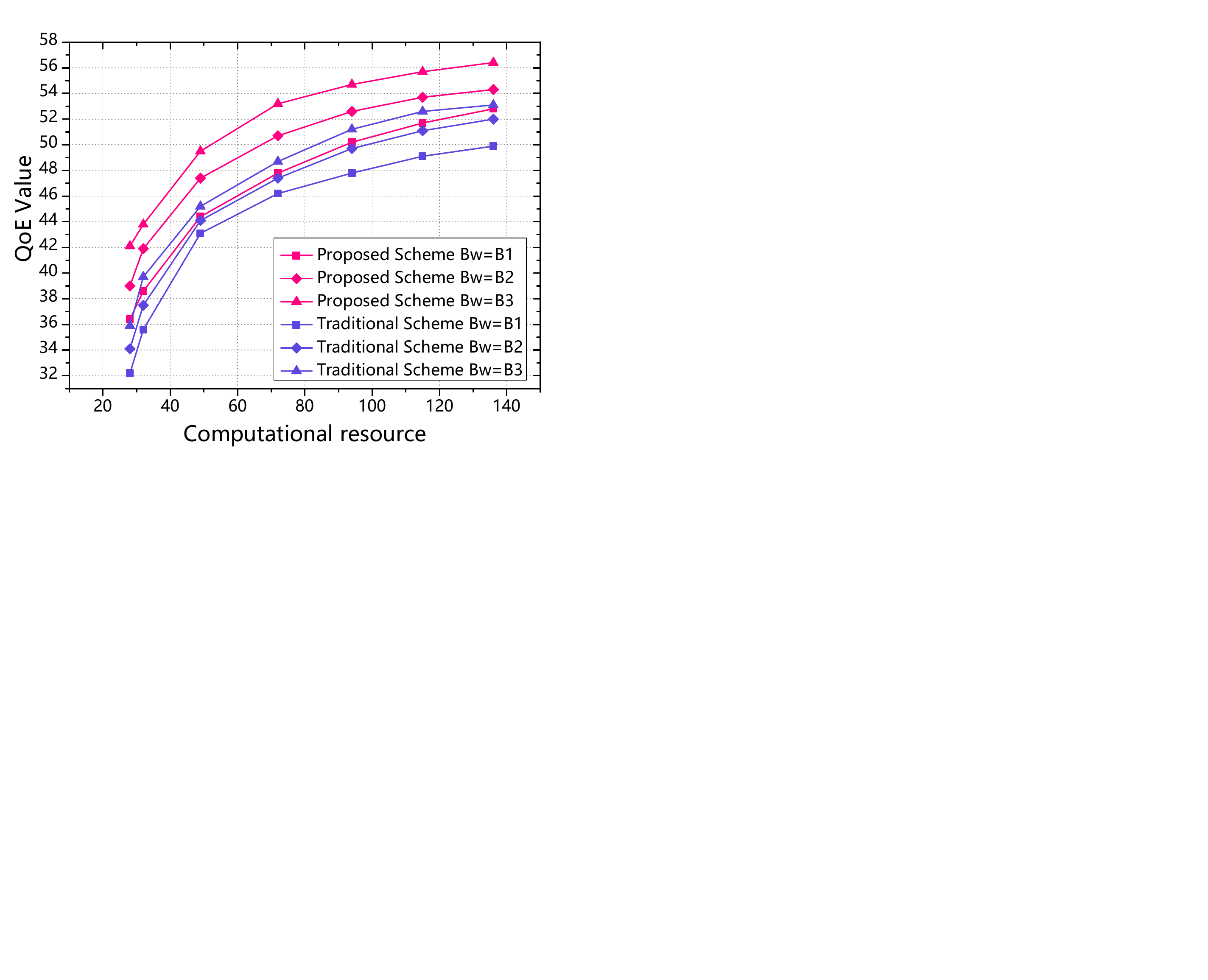}
    \caption{Maximum QoE comparison}
    \label{qoe_compre}
\end{figure}

Fig.\ref{qoe_compre} is a comparison of QoE values between our scheme and the traditional scheme. We can see that since our transmission scheme makes full use of communication resources and computational resources, it can maximize the quality levels of the tiles, and can obtain a greater QoE value. We can also observe that when the computing resources are scare, which is typical for mobile user playback equipment, the QoE value is significantly higher than the traditional solution.

\section{CONCLUSION} \label{sec_conculsion}

In this paper, we proposed a communication and computational resource allocation scheme for QoE-driven point cloud video streaming. In particular, with the goal to maximize the defined QoE by selecting
proper quality levels (uncompressed tiles at different quality levels are also considered) for each partitioned point cloud video tile, we formulated this into an optimization problem under the limited communication and computational resources constraints and proposed a scheme to solve it. Extensive simulations were conducted and the simulation results showed the superior performance of the proposed scheme over the existing schemes.

%
\bibliographystyle{IEEEtran}
\bibliography{reference}
\ifCLASSOPTIONcaptionsoff
  \newpage
\fi

\end{document}